\documentclass[sigconf,acmthm=false]{acmart}
\usepackage{mathtools}
\usepackage{thmtools,thm-restate}
\usepackage{enumitem}
\usepackage{hyperref}
\usepackage[ruled,vlined,linesnumbered]{algorithm2e}
\usepackage{tikz}
\usepackage[textsize=tiny,disable]{todonotes}

\usepackage{xcolor}     
\usepackage[normalem]{ulem} 

\makeatletter
\@ifundefined{newcounteralias}{}{%
  \renewcommand\thmt@autorefsetup{%
    \@xa\def\csname\thmt@envname autorefname\@xa\endcsname\@xa{\thmt@thmname}}%
}
\makeatother

\newtheorem{theorem}{Theorem}[section]
\newtheorem{lemma}[theorem]{Lemma}
\newtheorem{claim}[theorem]{Claim}
\newtheorem{proposition}[theorem]{Proposition}

\theoremstyle{definition}
\newtheorem{definition}[theorem]{Definition}

\definecolor{layerone}{RGB}{86,180,233}
\definecolor{layertwo}{RGB}{230,159,0}
\definecolor{layerthree}{RGB}{0,158,115}
\definecolor{layerfour}{RGB}{204,121,167}
\definecolor{layerfive}{RGB}{213,94,0}

\newcommand{\E}{\mathbb{E}}
\newcommand{\Prb}{\mathbb{P}}

\newcommand{\defn}[1]{\textbf{\emph{\boldmath #1}}}
\newcommand{\ind}{\mathbf{1}}
\newcommand{\prot}[1]{\textnormal{\textsc{#1}}}

\newif\ifcomments
\commentsfalse

\newcommand{\sidetodo}[3]{%
  \ifx l#1%
    {\reversemarginpar\todo[color=#2]{#3}}%
  \else
    \todo[color=#2]{#3}%
  \fi
}

\newcommand{\pkm}[2][r]{\sidetodo{#1}{violet!20}{Pritom: #2}}

\newif\ifdraft
\drafttrue

\newbox\wipbox
  {\ifdraft\color{carnelian}\else\setbox\wipbox=\vbox\bgroup\fi}%
  {\ifdraft\else\egroup\fi}

\setcopyright{none}
\renewcommand\footnotetextcopyrightpermission[1]{}
\hypersetup{hypertexnames=false}
\makeatletter
\AtBeginDocument{%
  \fancypagestyle{submission}{%
    \fancyhf{}%
    \fancyhead[L]{\ACM@linecountL}%
    \fancyhead[R]{\ACM@linecountR}%
    \fancyfoot[C]{\footnotesize\thepage}}%
  \pagestyle{submission}}
\makeatother

\begin{document}

\title{Contention Resolution Against an Adversarial Scheduler}
\author{Michael A. Bender}
\affiliation{%
  \institution{Stony Brook University}
  \city{Stony Brook}
  \country{USA}
}

\author{Mart\'{\i}n Farach-Colton}
\affiliation{%
  \institution{New York University}
  \city{New York}
  \country{USA}
}

\author{Andrew Krapivin}
\affiliation{%
  \institution{Carnegie Mellon University}
  \city{Pittsburgh}
  \country{USA}
}

\author{Pritom Kundu}
\affiliation{%
  \institution{Stony Brook University}
  \city{Stony Brook}
  \country{USA}
}

\author{John Kuszmaul}
\affiliation{%
  \institution{Massachusetts Institute of Technology}
  \city{Cambridge}
  \country{USA}
}

\author{Margaret Kuszmaul}
\affiliation{%
  \institution{Northeastern University}
  \city{Boston}
  \country{USA}
}

\author{William Kuszmaul}
\affiliation{%
  \institution{Carnegie Mellon University}
  \city{Pittsburgh}
  \country{USA}
}

\author{Jingxun Liang}
\affiliation{%
  \institution{Carnegie Mellon University}
  \city{Pittsburgh}
  \country{USA}
}

\begin{abstract}
In the classical contention-resolution problem, a set of players share a broadcast
channel, and a player succeeds in a given time step if it is the only player to
broadcast during that step. Implicit in the classical problem is the assumption that
every player is free to act on every step: once a player arrives, it participates
until it succeeds.

We study what happens when that assumption is removed. In our model, an adversarial
\emph{scheduler} decides, on each step, which subset of the players gets scheduled
to participate in the step. An unscheduled player simply sleeps through the step. We measure a
protocol by its \emph{efficiency}: the number of players divided by the expected
number of steps in which at least one player was active.

Our first result says that, so long as the players are given access to a global clock and 
to shared randomness, they can achieve efficiency $\Omega(1/\log n)$. 
The players are able to achieve this bound without requiring any channel feedback, and 
even if the adversarial scheduler is adaptive. We also
give a matching lower bound of $O(1 / \log n)$, which holds even with channel feedback, and even for an
oblivious scheduler.

Our upper bounds all require the players to have access to a shared global clock and to shared randomness. 
Our final result says that this type of implicit shared state is necessary. If the players have neither a global clock nor shared randomness,
then no protocol can achieve efficiency better than $O(\log n / n)$; and this holds even if the other aspects of the problem 
are designed to be as favorable for the players as possible (the channel gives ternary feedback and the adversary is oblivious). 

\end{abstract}

\pagenumbering{gobble}
\maketitle
\thispagestyle{submission}
\clearpage
\pagenumbering{arabic}

\section{Introduction}
\label{sec:intro}

In the \defn{contention resolution problem} there are $n$ \defn{players} that need to coordinate temporary and exclusive access to a shared resource.
Commonly, we talk about contention resolution in language borrowed from networking, where the shared resource is a communication channel, and each player has a message that it must send on that channel. This is the terminology that we adopt in this paper, although contention resolution is used in a wide range of settings, including WiFi~\cite{IEEE-802.11-1999,xiao:performance} and cellular networks~\cite{LTE}, shared-memory architectures~\cite{Ben-DavidB17,BenderBlFa26}, concurrency control~\cite{80120,10.1145/173682.165164}, distributed computing~\cite{luby1986simple}, email retransmission~\cite{Bernstein98,CostalesAl02}, congestion control in TCP~\cite{jacobson:congestion}, and cloud computing~\cite{amazon:error-retries,google:best-practices}.

Formally, the shared channel is called a \defn{multiple-access channel}~\cite{Goldberg-notes-2000,ChlamtacKu85}.
Each player has a message it wishes to transmit on the channel.
Time is divided into \defn{steps}, each of which is sized to fit a single message. A player who tries to send their message on a given step is said to \defn{broadcast} on that step. A broadcast results in a successful message transmission if \emph{no other players} broadcast on the same step~\cite{AbramsonKu73,MetcalfeBo76,Goldberg-notes-2000}.
If two or more players try to send on the channel in the same step, there is a \defn{collision}, and none of the message transmissions succeed. Once a player successfully transmits its message, that player leaves the system.

The goal is for players to make use of the channel with as few wasted steps as possible. Formally, this means that the players wish to maximize the \defn{efficiency} $n/\E[L]$, where $L$ is the total number of steps that it takes for all players to succeed, ignoring steps in which there are no players active in the system~\cite{BenderFiGi19} (we will come back to this distinction shortly). Given this goal, contention resolution becomes a problem of cooperation: how can the players coordinate, either implicitly or explicitly, to avoid either \emph{overusing} the channel (resulting in collisions) or \emph{underusing} it (resulting in silence). 

Contention resolution has a rich literature dating back to work in the 1970s on ALOHA satellite scheduling~\cite{ALOHAnet} and binary exponential backoff~\cite{MetcalfeBo76}. Much of the work on the problem has focused on how the following two aspects of the model affect the players' optimal strategies:
\begin{enumerate}
\item \textbf{How players enter the system:} Do the players all arrive at once, or arrive over time according to an adversary or a stochastic process? If the arrivals are adversarial, is the adversary oblivious or adaptive?
\item \textbf{What feedback players hear:} Do players who choose not to broadcast still learn whether a given step was a success or a failure? On failure, do they learn whether there was a collision or silence? Or do they learn nothing at all? 
\end{enumerate}
Depending on the answers to these questions, one obtains results of very different flavors. In settings where players receive feedback~\cite{BenderKoPe18,BenderFiGi19,ChangJiPe19} (even limited feedback~\cite{BenderKoKu20,chen2021tight}), they can use this feedback to coordinate with one another in order to achieve efficiency $\Omega(1)$, even in the face of adversarial arrivals~\cite{BenderKoPe18,BenderFiGi19,ChangJiPe19,BenderKoKu20,chen2021tight}. In settings where players do not receive feedback, they can still hope to use cleverly designed backoff/back-on protocols to achieve good efficiency~\cite{hastad:analysis,greenberg1985randomized,gereb1992efficient,BenderFaHe05}: if the players all enter the system at once, they can achieve efficiency $\Omega(1)$~\cite{greenberg1985randomized,gereb1992efficient,BenderFaHe04,BenderFaHe05}; if the players enter the system over time, then so long as the arrivals are sufficiently spread out~\cite{BenderFaHe05} (and determined by an oblivious adversary), the players can achieve efficiency $\Omega(1/\log n)$. There has also been work on many other variations of the problem, e.g., in which the goal is to minimize energy consumption rather than efficiency~\cite{BenderKoPe18,BenderFiGi19,HradovichKlKo20,BenderKoKu20,DBLP:conf/icdcs/MarcoKS21,DeMarcoKoSt22,DeMarcoKoSt22a,DBLP:conf/spaa/JiangZ22,BenderFiGi24,DBLP:journals/siamcomp/BenderFGKY25,DBLP:conf/spaa/XieZ25}, in which the players can communicate explicitly rather than through feedback~\cite{BenderKoPe18,DBLP:conf/spaa/AgrawalBFGY20,DeMarcoKoSt22a,BenderGiKu22}, and in which an adversary can \emph{jam} some steps, causing broadcasts in those steps to automatically fail~\cite{DBLP:conf/podc/AwerbuchRS08,DBLP:conf/sirocco/AnantharamuCKR11,DBLP:conf/icdcs/RichaSSZ11,DBLP:conf/podc/RichaSSZ12,DBLP:journals/talg/AwerbuchRSSZ14,BenderFiGi19,DBLP:journals/jcss/AnantharamuCKR19,ChangJiPe19,BenderKoKu20,chen2021tight,DBLP:conf/spaa/JiangZ22,BenderFiGi24,DBLP:journals/siamcomp/BenderFGKY25}, etc. See \cite{BanicescuChGi25} for a survey on the topic.


Yet one important source of uncertainty is largely absent from these models: whether a player is able to participate at all at a given time step. Standard contention-resolution models typically assume that, once a player arrives, it progresses through every time step and always has an opportunity to broadcast or observe the channel. 
The uncertainty of when a player gets to take steps, however, is a fundamental concern in asynchronous systems.
For example, players may represent system processes, which can be delayed by preemptive scheduling or page faults, or individual devices, which may sleep to conserve energy, become temporarily unavailable, or need to swap between multiple tasks~\cite{scherer2005advanced,fich2005obstruction,Ben-DavidB17}. In all of these cases, different players may miss different parts of the execution.

This raises the following natural question: What happens if we allow an adversary to determine not just \emph{when} each player arrives, but \emph{which} time steps that player is scheduled for? This is to say that a (possibly adaptive) scheduler gets to choose a subset of the players to be \defn{scheduled} in a given time step. Players that are scheduled experience the step like a normal time step. Players that are unscheduled do not get to participate at all -- they cannot broadcast and they cannot receive feedback on that step.

At first glance, the power of the adversary seems to make efficient contention resolution nearly impossible: players may miss arbitrarily large portions of the execution and thus have radically different views of the state of the system. The main purpose of this paper is to show that this is not the case: even against an adaptive scheduler, efficient contention resolution remains possible. At the same time, as we will see, the presence of the scheduler \emph{does} significantly change the problem. The algorithmic techniques that the players should use become entirely different; models that are normally very different (depending on feedback or strength of adversary) collapse to have the same asymptotic guarantees; and modeling distinctions that may seem less significant \emph{a priori} (do the players have access to a global clock and shared randomness) end up playing make-or-break roles for whether the players can succeed.

\paragraph{Results.} The first result of this paper is a surprisingly simple upper bound. We show that, so long as the players have access to a few basic resources -- a global clock to tell them the time and a shared string of random bits -- they can achieve an efficiency of $\Omega(1 / \log n)$. Moreover, this result holds even for the most adversarial version of the problem: against an adaptive scheduling adversary, and without any type of feedback from the channel. 

\begin{restatable}{theorem}{variantone}
\label{thm:variant1}
Suppose the players have a global clock, shared randomness, and knowledge of $n$. Against any adaptive scheduler, and without any channel feedback, there exists a protocol (the \prot{Guess-a-Scale} protocol) that achieves
\[
  \E[L] \;\le\; 32\big(\lfloor \log n \rfloor + 1\big)\, n,
\]
and hence achieves efficiency $\Omega(1/\log n)$.
\end{restatable}

The basic idea behind Theorem \ref{thm:variant1} is that, rather than treating contention resolution as a problem about back-off/back-on protocols, the players treat it more like a guessing game. On each step, the players use their shared randomness to make a shared guess as to roughly how many players they think are scheduled during that step. The players then use this shared guess to determine with what probability each player should (independently) broadcast. The result is a completely memoryless feedback-free algorithm that still achieves relatively high efficiency.

Note that Theorem \ref{thm:variant1} is interesting even for the classical version of the problem where the scheduler determines player arrivals and nothing else. In this version of the problem, and in the setting of Theorem \ref{thm:variant1} where players receive no feedback, it was known \cite{BenderFaHe05} how to achieve efficiency $\Omega(1 / \log n)$ only with a substantially limited adversary (one that is oblivious and that is limited in how quickly it can enter players into the system). Theorem \ref{thm:variant1} says that, with a global clock and shared randomness, this limited adversary can be replaced with an arbitrary fully adaptive one, and even one that is permitted to schedule players' entire lives rather than just their arrivals. 

The second result of the paper is a lower bound showing that Theorem \ref{thm:variant1} is optimal.

\begin{restatable}{theorem}{lbclock}
\label{thm:lb-clock}
There exists an oblivious scheduler against which every player protocol satisfies $\E[L]=\Omega(n\log n)$ and hence has efficiency $O(1/\log n)$. This holds even when the players have a global clock, shared randomness, ternary feedback, and knowledge of $n$. In fact, it holds even if players are given complete information about the past (i.e., for each step, which players were scheduled and which ones broadcast).
\end{restatable}

Note that Theorem \ref{thm:lb-clock} applies even in the setting that is \emph{most generous} to the players. It restricts the scheduler to be oblivious, and it gives players access not just to per-step feedback but to all information about the past. In the classical version of the problem, these distinctions determine the optimal efficiency, but in the setting of Theorem \ref{thm:lb-clock}, they make no difference -- no matter the strength of the adversary or the type of feedback the players receive, the optimal efficiency remains $\Theta(1/\log n)$. 

In addition to the theorems above, we also consider several extensions of the model that make the problem more difficult for the players. 

In Section \ref{sec:coarse}, we consider what happens if the global clock that the players access gives them only a coarse-grained measurement of time, reporting $\lceil t / k \rceil$, where $t$ is the true time, and $k$ is some granularity parameter. 
If the scheduler is adaptive, this tanks the optimal efficiency for the players, transforming it from $\Theta(1/\log n)$ to $\Theta(1/(k\log n))$ (see Appendix \ref{app:adaptive-coarse}). On the other hand, if the scheduler is oblivious, then something surprising happens: so long as $k$ is not too large ($k \le O(\log n)$), the players can still achieve efficiency $\Theta(1 / \log n)$, independent of $k$. Thus, whereas for a $1$-granular clock the adaptiveness of the scheduler does not matter, for a $k$-granular clock it does. 
This result, which is formalized in Theorem \ref{thm:variant2}, requires a number of additional algorithmic/analytical ideas beyond Theorem \ref{thm:variant1}. 

We also consider in Section \ref{sec:unknownn} a second extension to the model in which the players do not know $n$ in advance. We show that, in this setting, the players can still achieve efficiency very close to $\Theta(1 / \log n)$ (Theorem \ref{thm:variant3}).

Finally, in Section \ref{sec:shared}, we examine more closely the two resources that appear essential for our algorithms: access to a shared global clock and to shared random bits. These resources can both be viewed as examples of \emph{implicit shared state}, state that allows the players to coordinate without explicitly communicating. Our final result says that such state is fundamentally necessary for good efficiency to be achieved.

\begin{restatable}{theorem}{lbnoclock}
\label{thm:lb-noclock}
Suppose that players have neither a global clock nor shared randomness. Then for every player protocol, there is an oblivious scheduler under which the expected number of live steps satisfies $\E[L] = \Omega(n^2 / \log n)$; the protocol's efficiency is therefore $O(\log n / n)$. This holds even if the players know $n$ and receive ternary feedback on the steps where they are scheduled. 
\end{restatable}

Note that, if players always broadcast with probability exactly $1/n$, they can trivially achieve efficiency $\Omega(1/n)$ (Proposition \ref{prop:trivial}). Thus, Theorem \ref{thm:lb-noclock} is very close to being the most aggressive type of lower bound one could ask for. It says that without access to shared implicit state, and even if everything else is designed to be in the players' favor, they cannot hope to do significantly better than the trivial protocol. 

We remark that the role of a global clock has also been studied in the classical setting, where the scheduler controls arrivals only. Notably, for the objective function of minimizing \emph{latency}, \cite{CaiChDu26} shows that there is a separation between models with and without global clocks. Theorem \ref{thm:lb-noclock} can be viewed as offering an analogous separation in our model, but for the case where the global clock is considered together with shared randomness. Although aesthetically similar, it should be noted that the results in \cite{CaiChDu26} are technically very different from those studied here.

\section{Preliminaries}
\label{sec:prelims}

We begin by describing the model that will be used throughout the paper. Throughout, $\log$ denotes the base-2 logarithm and $\ln$ the natural logarithm.

\paragraph{Players, time steps, and the channel.} There are $n$ \defn{players}, each with a single packet to send over a shared \defn{channel}. Time is divided into discrete \defn{steps} $t = 1, 2, 3, \ldots$. On each step, a player either \defn{broadcasts} or remains silent. A broadcast on step $t$ \defn{succeeds} if and only if exactly one player broadcasts on step $t$.
In particular, if two or more players broadcast on the same step, their transmissions \defn{collide}, and all of the broadcasts fail. Once a player issues a successful broadcast, that player leaves the system and cannot broadcast anymore; we refer to the players that have not yet left as \defn{present}.

Each player is aware of its own past actions and private random bits. Depending on the variation of the model being considered, it may also be given access to the parameter $n$, a global clock, shared randomness, or feedback from the channel (defined below). We will specify, for each theorem that we prove, which of these features the players are or are not given.

\paragraph{The scheduler.} 
In our model, only a subset of players is allowed to participate in each step, which is chosen by an adversary. Formally, we model the adversary as a \defn{scheduler} which determines on each step $t$ a set $S_t$ of players that are permitted to broadcast on that step. The players in $S_t$ are \defn{scheduled} on step $t$, while every other player is \defn{unscheduled}. We say that a player is \defn{active} on step $t$ if it is both scheduled and present on step $t$, and that it is \defn{inactive} otherwise (i.e., if it is unscheduled or has already left the system). Only active players can broadcast; in particular, a scheduled player that is no longer present cannot.
Each player is only aware of the steps in which it is scheduled. And, unless the player is given access to a clock, it does not know any information about the steps in which it was unscheduled.

Our results will consider two types of schedulers, oblivious and adaptive. Both schedulers know the protocol that the players are following, but only an adaptive scheduler may respond adaptively to the players' behaviors. Formally, an \defn{oblivious} scheduler fixes the entire sequence $S_1, S_2, S_3, \ldots$ before the first time step, independently of the players' random choices. In contrast, an \defn{adaptive} scheduler may choose $S_t$ based on the players' past behavior, namely, which players issued broadcasts on all of steps $1, 2, \ldots, t - 1$. Note that the adaptive scheduler still must commit to $S_t$ before observing which players choose to broadcast on step $t$ (it cannot, for example, look at which players will broadcast, and define $S_t$ to be the complement set). Neither scheduler is able to directly see the random bits used by the players. 

We remark that the schedulers described above are significantly more powerful than schedulers that have been considered in past work on contention resolution. Past works have typically either considered no scheduler, or considered a scheduler whose only job is to pick \emph{arrival times} for each player~\cite{BenderFaHe05,DeMarcoSt17,CaiChDu26}. The scheduler in our model can not only choose each player's arrival time implicitly, by leaving that player unscheduled on all earlier steps, but also capture other kinds of asynchronous behavior, such as processes that are preempted by the operating system or devices that run at different speeds.

It is also worth noting that the schedulers we study differ significantly from the notion of a \emph{jamming adversary} \cite{DBLP:conf/podc/AwerbuchRS08,chen2021tight,DBLP:conf/spaa/JiangZ22,DBLP:journals/siamcomp/BenderFGKY25}, which can interfere with the channel on selected steps and prevent any broadcast from succeeding, even if exactly one player broadcasts. A jamming adversary is not an asynchronous scheduler: regardless of which steps are jammed, all players participate in all steps between when they arrive and successfully transmit. 

\paragraph{Live steps and efficiency.} The time cost of a protocol (which the players try to minimize and the scheduler tries to maximize) is the number of \defn{live steps} over the entire execution, where a step $t$ is considered \defn{live} if there is at least one player $i \in S_t$ that has not yet succeeded (i.e., there is at least one active player). 
Steps on which no player is active are not considered to be live. Let
\[
  L \;=\; \#\{ t : \text{step } t \text{ is live} \}
\]
denote the total number of live steps over the entire execution. Note that, once the player protocol has been selected, and for a given (potentially randomized) scheduler, $L$ is a random variable whose randomness comes from both the players and the scheduler.

The \defn{efficiency} (also known as \defn{implicit throughput}) of a protocol against a given scheduler is $n / \E[L]$. The (worst-case) efficiency of a protocol is its worst-case efficiency over all schedulers in the model under consideration. An efficiency of, say, $1/\log n$ means that the players convert one out of every $\log n$ broadcast opportunities into a success, on average. 
\pkm{Michael says: In distributed computing, generally the adversary is not really seen as having an algorithms, instead just the adversary just does adversarial stuff. We describe the adversary as a scheduler with a randomized algorithm which may bother people in distributed computing. Even if we are correct, we still should address this as this will bother people.}
\medskip

As discussed above, an active player's decision whether to broadcast depends on the information available to it. We vary this information across our theorems along four dimensions: access to the parameter $n$, access to a clock, shared randomness, and channel feedback. These are discussed in more detail below.

\paragraph{Access to \boldmath $n$.} For most of the results in this paper, we will assume the players know $n$ in advance. We will see, however, that removing this assumption only slightly affects the efficiency that the players can achieve (Section \ref{sec:unknownn}).

\paragraph{Access to a clock.} 
In every setting, a player can maintain a private counter recording the number of steps on which it has been scheduled. The clock assumptions specify what additional information about global time is available. With a \defn{global clock}, each scheduled player can read the current step number $t$. With a \defn{$k$-granular clock}, it can instead read $\lceil t/k \rceil$. Thus, a global clock is a $1$-granular clock, and a $k$-granular clock for a large $k$ is closer to no clock at all.

\paragraph{Shared randomness.} 
In every setting, each player has access to its own private random bits, independent of those of the other players. With \defn{shared randomness}, the players additionally have access to a common infinite string of random bits, independent of their private randomness, fixed before the execution begins and hidden from the adversary.

\paragraph{Feedback.} 
In every setting, a player learns whether its own broadcast succeeds, so that it can leave the system upon success.  With \defn{no feedback}, players receive no additional information from the channel. With \defn{ternary feedback} (also known as \defn{collision detection}), scheduled players can listen on the channel in each step, and learn whether it was \defn{silent} (no players broadcast), \defn{successful} (only one player broadcast) or \defn{noisy} (2+ players broadcast). Note that, in some settings, one also considers binary feedback (which distinguishes successes and failures without distinguishing silent versus noisy steps). In this paper, however, all upper bounds will apply with no feedback and all lower bounds will apply with full ternary feedback, so there will be no need to consider any intermediate notions.

\section{Basic Tools}

In this section, we present three simple lemmas that will be used throughout the rest of the paper.

\paragraph{The one-success formula.} 
The first lemma relates the probability of a successful transmission to the number of broadcasting candidates and their broadcast probability.

\begin{lemma}[The one-success formula]
\label{lem:onesuccess}
Suppose $J \ge 1$ players are active and each broadcasts independently with probability $p$. Then the probability that the step is a success is exactly $J p (1-p)^{J-1}$. Consequently:
\begin{enumerate}[label=(\alph*)]
  \item If $J = 1$ and $p = 1$, the step succeeds with probability $1$.
  \item If $p \le 1/2$ and $\tfrac12 \le Jp \le 2$, the step succeeds with probability at least $1/32$.
\end{enumerate}
\end{lemma}

\begin{proof}
The formula is immediate: there are $J$ ways to choose the lone broadcaster, and the step is a success exactly when that player broadcasts and the other $J-1$ do not. Part (a) is the case $J=1$, $p=1$. For part (b), recall that $(1-p)^{1/p} \ge 1/4$ for all $p \in (0, 1/2]$.\footnote{The function $(1-p)^{1/p}$ increases from $1/4$ at $p=1/2$ toward $1/e$ as $p \to 0$.} Writing $(1-p)^{J-1} = \big((1-p)^{1/p}\big)^{p(J-1)}$ and using $p(J-1) < Jp \le 2$, we get $(1-p)^{J-1} \ge 4^{-2} = 1/16$. Multiplying by $Jp \ge 1/2$ gives $1/32$.
\end{proof}

Lemma~\ref{lem:onesuccess} says that the broadcast probability only has to be right to within a constant factor. If $p$ is within a factor of two of $1/J$, a constant fraction of such steps succeed. Note that if $p$ is off by a large factor in either direction, the step is nearly worthless: with small $p$ there are no broadcasts, and with large $p$ there are collisions.

\paragraph{The accounting lemmas.} The next two lemmas allow us to obtain bounds on $\E[L]$ by proving bounds on how each individual step behaves. One lemma is for upper bounds while the other is for lower bounds. 

For $t \ge 0$, let the \defn{history} $H_t$ be the random variable keeping track of, for every step $1, \ldots, t$, which players were scheduled, which broadcast, and whether the step was a success. The first accounting lemma says that, if every live step has probability at least $q$ of succeeding, then the efficiency is at least $q$. 

\begin{lemma}[Accounting lemma for upper bounds]
\label{lem:accounting-ub}
Let $q > 0$, and suppose that
\[
  \Prb[\,\text{step } t \text{ succeeds} \mid H_{t-1} = h,\ S_t = S\,] \;\ge\; q
\]
for every step $t$, every possible history $h$ for the first $t - 1$ steps, and every set $S$ of scheduled players containing a player that is present according to $h$. Then $\E[L] \le n/q$, so the efficiency is at least $q$.
\end{lemma}
\begin{proof}
By assumption, $\Pr[\text{step }t\text{ succeeds} \mid \text{ step }t\text{ live}] \ge q$. Since there are at most $n$ successes over all time,
\begin{align*}
n & \ge \sum_t \Pr[\text{step }t\text{ succeeds}]  \\
& =  \sum_t \Pr[\text{step }t\text{ succeeds} \mid \text{step }t \text{ live}] \cdot \Pr[\text{step }t \text{ live}] \\
& \ge q \cdot \sum_t \Pr[\text{step }t \text{ live}] \\
& = q \cdot \E[L],
\end{align*}
as desired. 
\end{proof}

Finally, the second accounting lemma gives a simple condition under which we can conclude that the efficiency is small.
\begin{lemma}[Accounting lemma for lower bounds]
\label{lem:accounting-lb}
Let $q, c > 0$. Suppose that
\[
  \Prb[\,\text{step } t \text{ succeeds} \mid H_{t-1} = h\,] \;\le\; q
\]
and
\[
  \Prb[\,\text{step } t \text{ is live} \mid H_{t-1} = h\,] \;\ge\; c
\]
for every step $t$ and every possible history $h$ for the first $t - 1$ steps in which at least $n/2$ players are present. Then $\E[L] \ge cn/(2q)$, so the efficiency is at most $2q/c$.
\end{lemma}
\begin{proof}
Let $A_t$ be the event that at least $n/2$ players are present at the start of step $t$; it is determined by $H_{t-1}$. Discarding the histories in which $A_t$ fails and using the second hypothesis on the rest,
\[
  \E[L] \;=\; \sum_{t \ge 1} \Prb[\,\text{step } t \text{ is live}\,]
  \;\ge\; c \sum_{t \ge 1}\Prb[A_t],
\]
so it suffices to show that $\sum_{t \ge 1} \Prb[A_t] \ge n/(2q)$. If, with positive probability, $A_t$ occurs for all $t$, then the sum is infinite. Otherwise the number of present players falls below $n/2$ with probability $1$, eventually, so at least $n/2$ successes occur on steps where $A_t$ occurs. In this case, defining $Z_t$ to be the event that step $t$ is a success, and using $\ind[A_t]$ to denote the indicator random variable for event $A_t$, we can apply the first hypothesis to get
\[
  \frac n2 \;\le\; \E\Big[\sum_{t \ge 1} Z_t \cdot \ind[A_t]\Big] \;\le\; q \sum_{t \ge 1} \Prb[A_t].
\]
Rearranging gives $\sum_{t \ge 1} \Prb[A_t] \ge n/(2q)$, as desired.
\end{proof}

\section{The Most Basic Setting: Fine-Grained Clock and Known \texorpdfstring{\boldmath $n$}{n}}
\label{sec:basic}
In this section, we consider what happens if the players are given access to both a global clock and to shared randomness and get to know $n$.
In this setting we show that the optimal efficiency is $\Theta(1 / \log n)$. Our upper bound holds even under a pessimal setup for the players (an adaptive scheduler and with no feedback), while our lower bound holds even under a pessimal setup for the scheduler (an oblivious scheduler and with ternary feedback). In fact, as we will see, the lower bound even holds in a slightly stronger setting in which players are given full access to \emph{all information} about the history of the system. 

\subsection{The \texorpdfstring{\prot{Guess-a-Scale}}{Guess-a-Scale} Protocol} Our upper bound will be achieved via the following simple protocol. Let
\[
  \mathcal{I} = \{0,1,\ldots,\lfloor \log n \rfloor\},
  \qquad |\mathcal{I}| = \lfloor \log n \rfloor + 1,
\]
and call each $i \in \mathcal{I}$ a \defn{scale}. Using their shared randomness, the players agree on an infinite sequence $i_1, i_2, \ldots$ of scales, where the $i_t$s are independent and uniformly random scales $i_t \in \mathcal{I}$. Then, on each step $t$, the players that are active each broadcast independently (using their private randomness) with probability $1 / 2^{i_t}$. We refer to this algorithm as the \prot{Guess-a-Scale} protocol. 

To get a sense of why this works, consider some step $t$, let $J$ be the number of players that are active during that step, and suppose $J \ge 1$. To obtain a constant success probability, the players would like to broadcast with probability about $1/J$, but they do not know $J$. They do, however, know that $1 \le J \le n$, so Lemma~\ref{lem:onesuccess} guarantees that one of the $|\mathcal{I}| = \Theta(\log n)$ scales yields a suitable broadcast probability. By selecting such a scale $i_t$ at random, and by using their shared randomness to ensure that \emph{all of the players use the same $i_t$}, the players have a $\Theta(1 / \log n)$ chance of selecting a good broadcast probability. This intuition is formalized in the following theorem. 

\variantone*

\begin{proof}
Fix a step $t$, and condition on the history of the first $t-1$ steps taking some particular value $H_{t-1} = h$ and on the scheduler scheduling some particular set $S_t = S$ on step $t$. This fixes the number $J$ of players that are active. The current scale $i_t$ is independent of everything we have conditioned on and is uniformly random from $\mathcal{I}$.

Suppose that step $t$ is live, meaning that $1 \le J$, and let $i^\star=\lfloor\log J\rfloor \in \mathcal{I}$. With probability $1/|\mathcal{I}|$, the players choose $i_t=i^\star$. Conditioned on this event, their broadcast probability $p=2^{-i^\star}$ satisfies $1 \le Jp < 2$. If $i^\star = 0$, then $J = 1$ and $p = 1$, so by Lemma~\ref{lem:onesuccess}(a) the step succeeds with probability $1$. Otherwise $i^\star \ge 1$, so $p \le 1/2$, and by Lemma~\ref{lem:onesuccess}(b) the conditional probability of a successful broadcast is at least $1/32$. Thus, for any history $H_{t - 1} = h$ and set $S_t = S$ that result in step $t$ being live, we have
\[
  \Prb[\,\text{step } t \text{ succeeds} \mid H_{t-1} = h,\ S_t = S\,]
  \;\ge\; \frac{1}{32\,|\mathcal{I}|}.
\]
By Lemma~\ref{lem:accounting-ub} with $q = 1/(32|\mathcal{I}|)$, it follows that $\E[L] \le 32\,|\mathcal{I}|\,n$.
\end{proof}

It is worth pausing to appreciate how much this simple protocol achieves. The scheduler may choose arrival times adaptively, schedule any subset of players on any step, observe every success, and use the entire history to plan its next move. Each player, meanwhile, simply reads the clock and flips a coin: it neither tracks the other players nor receives channel feedback beyond notification of its own success. The clock and shared randomness let the players agree on a fresh scale that the scheduler cannot predict. This coordination alone is enough to achieve efficiency $\Omega(1 / \log n)$.

\subsection{A Matching Lower Bound}
\label{sec:lb-clock}

Next we establish a matching lower bound. The basic idea is that, now, the \emph{scheduler} chooses a random scale $j_t \in \mathcal{I}$ on each step, and schedules each player independently with probability $1 / 2^{j_t}$. Once the history and the players' current random choices are fixed, each present player has a fixed decision about whether to broadcast if scheduled. Randomizing the scheduling probability $1/2^{j_t}$ makes it unlikely that exactly one of these potential broadcasters is selected. We show that successes occur with probability $O(1/\log n)$ per step, while steps remain live with constant probability as long as at least half the players are present.

\lbclock*

\begin{proof}
We start by defining the oblivious scheduler and then analyzing its effect on any protocol. 

\medskip
\noindent\textbf{The scheduler.} Independently on each step $t$, the scheduler chooses a uniformly random scale $j_t \in \mathcal{I}$ and schedules each of the $n$ players independently with probability $2^{-j_t}$. These choices are all made at the beginning of time (so that the scheduler is oblivious) and therefore do not depend on which players are still present on a given step. The set $S_t$ of scheduled players may therefore contain both players which are present and players which are not.

\medskip
\noindent\textbf{Successes are rare.} Fix a step $t$, and condition on the history of the first $t-1$ steps taking some particular value $H_{t-1} = h$, \emph{and on all the randomness the players will use on step $t$}. This fixes each player's decision whether to broadcast at step $t$ if scheduled: the history $h$ records everything the players know about the past, and the conditioning fixes their current random choices. For each player $p$, define
\[
  X_p = \ind[\, p \text{ would broadcast on step } t \text{ if scheduled} \,].
\]
Under the conditioning above, each $X_p$ is a fixed value in $\{0,1\}$. Let
\[
  a = \#\{p : p \text{ is present and } X_p = 1\}.
\]
Thus, $a$ counts the present players that would broadcast if scheduled, and $a$ is also fixed under the conditioning. The only remaining randomness at step $t$ is the scheduler's scale $j_t$ and its scheduling choices.

We now bound the probability that step $t$ succeeds under the preceding conditioning. If $a=0$, success is impossible, so assume $a\ge1$.
Conditioned further on $j_t=j$, the number of broadcasters has distribution $\mathrm{Bin}(a,2^{-j})$. The one-success formula therefore gives success probability $a\cdot2^{-j}(1-2^{-j})^{a-1}$. Averaging over $j$ while retaining the conditioning on the history and the players' current random choices, we obtain
\begin{align}
  &\Prb[\,\text{step } t \text{ succeeds} \mid H_{t-1} = h,\, \text{players' randomness at step }t\,] \nonumber\\
  \;=\;& \frac{1}{|\mathcal{I}|}\sum_{j \in \mathcal{I}} a\,2^{-j}\big(1 - 2^{-j}\big)^{a-1}
  \;\le\; \frac{1}{|\mathcal{I}|}\sum_{j \ge 0} a\,2^{-j}\big(1 - 2^{-j}\big)^{a-1}. \label{eq:bound-on-success-probability-1}
\end{align}
We claim that
\begin{equation}
\label{eq:bound-on-success-probability}
  \sum_{j \ge 0} a\,2^{-j}\big(1 - 2^{-j}\big)^{a-1} \;\le\; 14
  \qquad \text{for every integer } a \ge 1.
\end{equation}
This is the heart of the argument: only a constant number of values of $j$ give a scheduling probability within a constant factor of $1/a$, and the contributions decay at least geometrically as $2^{-j}$ moves away from this range in either direction.

The case $a = 1$ is immediate, since the sum is then
\[
  \sum_{j \ge 0} 2^{-j} \;=\; 2.
\]
So assume that $a \ge 2$, and write $\lambda_j = a\,2^{-j}$ for the expected number of broadcasters when the scheduling probability is $2^{-j}$. The $j$-th term of the sum is $\lambda_j\big(1 - \lambda_j/a\big)^{a-1}$, and applying $1 - x \le e^{-x}$ with $x = \lambda_j / a$, followed by $(a-1)/a \ge 1/2$, gives
\[
  \lambda_j\Big(1 - \frac{\lambda_j}{a}\Big)^{a-1}
  \;\le\; \lambda_j e^{-\lambda_j (a-1)/a}
  \;\le\; \lambda_j e^{-\lambda_j/2}.
\]
It therefore suffices to bound $\sum_{j \ge 0} \lambda_j e^{-\lambda_j/2}$ by $14$, and we do this by splitting the values of $j$ according to whether $\lambda_j \le 2$.

Since $\lambda_{j+1} = \lambda_j / 2$, the values $\lambda_j$ that are at most $2$ form a geometric sequence of ratio $1/2$ whose largest element is at most $2$. The values of $j$ with $\lambda_j \le 2$ therefore contribute at most
\[
  \sum_{j \,:\, \lambda_j \le 2} \lambda_j e^{-\lambda_j/2}
  \;\le\; \sum_{j \,:\, \lambda_j \le 2} \lambda_j
  \;\le\; 2 + 1 + \frac12 + \frac14 + \cdots \;=\; 4.
\]
For the values of $j$ with $\lambda_j > 2$, let $v_j = \lfloor \lambda_j \rfloor \ge 2$. These integers are distinct, because $\lambda_j > 2$ implies $\lambda_{j+1} = \lambda_j / 2 < \lambda_j - 1$ and hence $v_{j+1} \le v_j - 1$. Moreover $\lambda e^{-\lambda/2}$ is decreasing for $\lambda \ge 2$, so each such term satisfies $\lambda_j e^{-\lambda_j/2} \le v_j e^{-v_j/2}$. Summing over the distinct integers $v_j$, these values of $j$ contribute at most
\[
  \sum_{j \,:\, \lambda_j > 2} \lambda_j e^{-\lambda_j/2}
  \;\le\; \sum_{v \ge 2} v e^{-v/2} \;<\; 10.
\]
Adding the two bounds gives $\sum_{j \ge 0} \lambda_j e^{-\lambda_j/2} \le 4 + 10 = 14$, which proves~\eqref{eq:bound-on-success-probability}. 

Plugging \eqref{eq:bound-on-success-probability} into \eqref{eq:bound-on-success-probability-1} gives
\begin{align*}
  &\Prb[\,\text{step } t \text{ succeeds} \mid H_{t-1} = h,\, \text{players' randomness at step }t\,] \\
  &\qquad\le\; \frac{14}{|\mathcal{I}|} \;=\; O\!\left(\frac{1}{\log n}\right).
\end{align*}
Therefore, averaging over the players' current random choices gives
\[
  \Prb[\,\text{step } t \text{ succeeds} \mid H_{t-1} = h\,] \;=\; O\!\left(\frac{1}{\log n}\right).
\]

\medskip
\noindent\textbf{Live steps are common.} Consider a possible history $h$ in which at least $n/2$ players are present at the start of step $t$. For every scale $j_t=j$, the probability that no present player is scheduled is at most $(1-2^{-j})^{\#\text{present players}}\le(1-1/n)^{n/2}\le e^{-1/2}$, because $2^{-j}\ge2^{-\lfloor\log n\rfloor}\ge1/n$. Consequently,
\[
  \Prb[\,\text{step } t \text{ is live} \mid H_{t-1} = h\,]
  \;\ge\; 1 - e^{-1/2} \;\ge\; 0.39.
\]

\medskip
\noindent\textbf{Combining the bounds.} Finally, applying Lemma~\ref{lem:accounting-lb}, with $q = O(1/\log n)$ and $c = 0.39$, gives
\[
  \E[L] \;\ge\; \frac{c\,n}{2q} \;=\;  \Omega(n \log n). \qedhere
\]
\end{proof}

\section{Extending to a Coarse-Grained Clock}
\label{sec:coarse}

So far, we have assumed a $1$-granular global clock, meaning that every player knows the \emph{exact} time step $t$ that is currently occurring. In this section we consider what happens if, instead, players are given access to a $k$-granular global clock for some $k > 1$. Throughout the rest of the section, we assume a $k$-granular global clock, and for each $w = 1, 2, 3, \ldots$ we refer to the block of steps $\{(w-1)k + 1, (w-1)k + 2, \ldots, wk\}$, which are exactly the steps on which the clock reads $w$, as the \defn{window} $w$.

Note that the analysis of the \prot{Guess-a-Scale} algorithm (from Section \ref{sec:basic}) trivially extends to this setting by considering only the first live step in each window (if such a step exists), and gives an efficiency of $\Omega(1 / (k \log n))$. We show in Appendix \ref{app:adaptive-coarse} that, if the scheduler is adaptive, then this efficiency is optimal. This means that, morally, each window is only able to contribute one ``potentially useful'' step.

Perhaps surprisingly, though, if we are against an \emph{oblivious} scheduler, we can do better. The main result of this section is to show that, against an oblivious scheduler, the players can achieve efficiency $\Omega(1 / \max(k, \log n))$ (Theorem~\ref{thm:variant2}). This means that, so long as $k$ is not too large ($k \in O(\log n)$), the efficiency \emph{remains} $\Omega(1 / \log n)$. 

\paragraph{Why \prot{Guess-a-Scale} breaks.} We begin by discussing why the \prot{Guess-a-Scale} algorithm from the previous section \emph{does not} get an $\Omega(1 / \log n)$-efficiency bound in this new setting. 

Suppose we run the \prot{Guess-a-Scale} algorithm, but with one scale $i_w$ per window $w$. Then, the scheduler can attack the algorithm by scheduling only a single player for all $k$ steps of a window. The players draw one scale for the whole window, so with probability $\approx 1 - O(\log k / \log n)$ that scale is far too small, the lone player stays silent for all $k$ steps, and the window burns $k$ live steps for nothing. On the rare windows where the scale is large, the window still only produces one success. Averaging over the scale, a window of this kind produces about $(\log k)/\log n$ successes at a cost of about $k$ live steps, for an efficiency of roughly $\frac{\log k}{k \log n}$. Compared with Theorem~\ref{thm:variant1}, we have lost a factor of nearly $k$. 

What the above scheduler demonstrates is the following general problem: If the same players are scheduled in every step of the window, and if the number of active players is some number $r \ll k$, then there is a substantial asymmetry between the cost of the window (the $k$ steps that the players are present for if they don't all succeed) versus the benefit of the window (the up to $r$ successes that could occur). This is the problem that we must solve if we wish to continue achieving good efficiency. 

\paragraph{The \prot{Two-Mode} protocol.} The failure above happens only when the number $J$ of active players is small. Notice, however, that when $J$ is small, say $J \le \log n$, there is a trivial protocol that does reasonably well: if every player just broadcasts independently with probability $1/\log n$, then, until the first success of the window, each such step succeeds with probability $\Omega(1/\log n)$. This suggests the following \prot{Two-Mode} protocol, which chooses on each window randomly between the \prot{Guess-a-Scale} approach versus the broadcast-with-probability-$1/\log n$ approach. 

\begin{definition}[The \prot{Two-Mode} protocol]
\label{def:twomode}
For each window $w$, the players use their shared randomness to draw an independent fair coin and an independent uniform scale $i_w \in \mathcal{I}$. On every step of window $w$, every active player broadcasts independently with probability
\[
  p_w = \begin{cases}
    2^{-i_w} & \text{if the coin came up heads} \quad \text{(\defn{guessing mode})},\\[2pt]
    1/\log n & \text{if the coin came up tails} \quad \text{(\defn{steady mode})}.
  \end{cases}
\]
\end{definition}

The following theorem shows that, if $k \in O(\log n)$, and if the scheduler is oblivious, then the above algorithm achieves efficiency $\Omega(1 / \log n)$. 

\begin{restatable}{theorem}{varianttwo}
\label{thm:variant2}
Suppose $n \ge 2$, and suppose the players have a $k$-granular clock, shared randomness, and knowledge of $n$. Then against an oblivious scheduler, and without requiring any feedback, the \prot{Two-Mode} protocol satisfies $\E[L] = O(n \max(k, \log n))$, and therefore achieves efficiency $\Omega(1/\max(k, \log n))$.
\end{restatable}

Before the proof, we establish some notation for a single window. Fix a window $w$, and condition on the entire history up to the moment the window begins. Because the scheduler is oblivious, the sets $S_t$ and the arrival times are fixed in advance, so at this moment we already know, for each step $t$ of the window, the quantity
\begin{align*}
  J_t^{0} \;=\; \#\big\{\, p \;:\;& p \text{ is scheduled on step } t, \\
  & \text{and } p \text{ is present at the start of window } w \,\big\}.
\end{align*}
Think of $J_t^0$ as the number of players who would be active on step $t$ if nobody were to succeed during the window. Let $J_t \le J_t^0$ be the \emph{actual} number of active players on step $t$. The two differ by exactly the number of successes that have already occurred during the window among the relevant players. Finally set
\[
  M \;=\; \#\{ t \in w : J_t^{0} \ge 1 \},
\]
the number of steps of the window that \emph{could} be live. Since players only leave, every live step of the window is one of these, so the number $L_w$ of live steps in the window satisfies $L_w \le M \le k$. Let $X_w$ denote the number of successes during the window.

The key to proving the theorem is to establish the following lemma, which says that a window always pays for itself at a rate of $1/\max(k, \log n)$.

\begin{lemma}[Window lemma]
\label{lem:window}
In the setting of Theorem~\ref{thm:variant2}, and conditioned on any history up to the start of window $w$,
\[
  \E[X_w] \;\ge\; \frac{M}{512 \max(k, \log n)}.
\]
\end{lemma}

\begin{proof}[Proof of Lemma~\ref{lem:window}]
Split the potentially-live steps of the window according to how crowded they are:
\begin{align*}
  M_{\mathrm{small}} &= \#\{t \in w : 1 \le J_t^0 \le \log n\}, \\
  M_{\mathrm{big}} &= \#\{t \in w : J_t^0 > \log n\},
\end{align*}
so that $M = M_{\mathrm{small}} + M_{\mathrm{big}}$. We show separately that $\E[X_w] \ge \frac{M_{\mathrm{small}}}{12 \max(k, \log n)}$ and that $\E[X_w] \ge \frac{M_{\mathrm{big}}}{256 \max(k, \log n)}$. Since at least one of $M_{\mathrm{big}}$ or $M_{\mathrm{small}}$ is at least $M/2$, the lemma follows. 

\medskip
\noindent\textbf{The uncrowded steps.} Condition on steady mode, which happens with probability $1/2$; then $p_w = 1/\log n$ on every step of the window. Consider the steps $t$ with $1 \le J_t^0 \le \log n$, and process them in chronological order. If no success has yet occurred during the window, then no player has left, so $J_t = J_t^0$, and by the one-success formula the probability that step $t$ succeeds is
\begin{align*}
  J_t^0 \cdot \frac{1}{\log n}\Big(1 - \frac{1}{\log n}\Big)^{J_t^0 - 1}
  &\;\ge\; \frac{1}{\log n} \Big(1 - \frac{1}{\log n}\Big)^{\log n - 1} \\
  &\;\ge\; \frac{1}{e \log n} \;\ge\; \frac{1}{3 \log n},
\end{align*}
using $J_t^0 \ge 1$ for the first factor, $J_t^0 \le \log n$ in the exponent, and $(1 - 1/x)^{x-1} \ge 1/e$ for $x \ge 1$. Chaining this over the $M_{\mathrm{small}}$ such steps (at each one, conditioned on there being no success so far in the window, we get a success with probability at least $1/(3 \log n)$) gives
\begin{align*}
  \Prb[\, X_w \ge 1 \mid \text{steady mode} \,]
  &\;\ge\; 1 - \Big(1 - \frac{1}{3 \log n}\Big)^{M_{\mathrm{small}}} \\
  &\;\ge\; 1 - e^{-M_{\mathrm{small}}/(3 \log n)} \\
  &\;\ge\; \frac12 \min\Big(1, \frac{M_{\mathrm{small}}}{3 \log n}\Big) \\
  &\;\ge\; \frac{M_{\mathrm{small}}}{6 \max(k, \log n)},
\end{align*}
where the third step uses $1 - e^{-x} \ge \min(x, 1)/2$ for $x \ge 0$, and the last step uses $M_{\mathrm{small}} \le k$. Multiplying by the probability $1/2$ of steady mode gives $\E[X_w] \ge M_{\mathrm{small}}/(12 \max(k, \log n))$, as claimed.

\medskip
\noindent\textbf{The crowded steps.} Now condition on guessing mode, which also occurs with probability $1/2$. For each scale $i \in \mathcal{I}$, let
\[
  m_i = \#\{ t \in w : J_t^0 > \log n \text{ and } 2^{i} \le J_t^0 < 2^{i+1} \},
\]
so that $\sum_{i} m_i = M_{\mathrm{big}}$. Fix a scale $i$ with $m_i \ge 1$ and condition further on $i_w = i$, an event that occurs with probability $1/|\mathcal{I}|$; the broadcast probability is then $p = 2^{-i}$ throughout the window. Call the $m_i$ steps counted by $m_i$ the \defn{designated} steps.

Consider a designated step $t$, and suppose that fewer than $(\log n)/2$ successes have occurred so far during the window. Then $J_t \ge J_t^0 - (\log n)/2 \ge J_t^0/2$, using $J_t^0 > \log n$. Combined with $J_t \le J_t^0 < 2^{i+1}$, this gives $\tfrac12 \le J_t \, p < 2$. Moreover $p \le 1/2$, since $J_t^0 > \log n \ge 1$ means $J_t^0 \ge 2$, which forces $i \ge 1$. So Lemma~\ref{lem:onesuccess}(b) applies: step $t$ succeeds with probability at least $1/32$.

In other words, walking through the designated steps in order (and conditioning on $i_w = i$), at each one either at least $(\log n)/2$ successes have already occurred in this window -- in which case $X_w \ge (\log n)/2$ and we are more than done -- or the step succeeds with conditional probability at least $1/32$. It follows that $X_w$ stochastically dominates $\min\big(B, (\log n)/2\big)$, where $B \sim \mathrm{Bin}(m_i, 1/32)$. To evaluate the expected value of this, note that $B \le m_i$ always, so $\min(B, \frac{\log n}{2}) \ge B \cdot \min\!\big(1, \tfrac{\log n}{2 m_i}\big)$, meaning that
\begin{align*}
  \E\Big[\min\Big(B, \frac{\log n}{2}\Big)\Big]
  &\;\ge\; \min\Big(1, \frac{\log n}{2m_i}\Big)\cdot \frac{m_i}{32} \\
  &\;=\; \frac{1}{32}\min\Big(m_i, \frac{\log n}{2}\Big) \\
  &\;\ge\; \frac{m_i \log n}{64 \max(k, \log n)},
\end{align*}
where the last step uses $m_i \le k$. Finally, removing the condition $i_w = i$, and observing that the events $\{i_w = i\}$ are disjoint $\frac{1}{|\mathcal{I}|}$-probability events across $i$,
\begin{align*}
  \E[X_w] &\;\ge\; \frac12 \sum_{i \in \mathcal{I}} \frac{1}{|\mathcal{I}|} \cdot
  \frac{m_i \log n}{64 \max(k, \log n)} \\
  &\;=\; \frac{\log n}{|\mathcal{I}|} \cdot \frac{M_{\mathrm{big}}}{128 \max(k, \log n)} \\
  &\;\ge\; \frac{M_{\mathrm{big}}}{256 \max(k, \log n)},
\end{align*}
where the last step uses $|\mathcal{I}| = \lfloor \log n \rfloor + 1 \le 2 \log n$ for $n \ge 2$.
\end{proof}

\begin{proof}[Proof of Theorem~\ref{thm:variant2}]
Since $M$ is determined at the start of the window, taking expectations in Lemma~\ref{lem:window} gives $\E[X_w] \ge \frac{\E[M_w]}{512 \max(k, \log n)}$ for every window $w$, where $M_w$ is the window's value of $M$. Summing over all windows and using $L_w \le M_w$ together with $\sum_w X_w \le n$,
\begin{align*}
  \E[L] &\;=\; \sum_w \E[L_w] \;\le\; \sum_w \E[M_w] \\
  &\;\le\; 512 \max(k, \log n) \sum_w \E[X_w] \\
  &\;\le\; 512\, n \max(k, \log n),
\end{align*}
as desired.
\end{proof}

We remark that Theorem \ref{thm:variant2} gives a nontrivial bound even for arbitrarily large window size $k$: when $k \gg \log n$, the efficiency is $\Omega(1/k)$. Whether this tradeoff is optimal for large $k$ remains an interesting open question.

\section{Extending to Unknown \texorpdfstring{\boldmath $n$}{n}}
\label{sec:unknownn}

Our final upper bound concerns the case where the players do not get to know $n$ in advance. We show that, in this setting, it is still possible for the players to achieve efficiency \emph{very close} to $1 / \log n$: for every constant integer $r \ge 1$ and every constant $\epsilon > 0$, there is a protocol whose efficiency is within a factor of $\ln^{(2)} n \cdots \ln^{(r)} n \cdot \big(\ln^{(r+1)} n\big)^{1+\epsilon}$ of $1/\ln n$, where $\ln^{(k)}$ denotes the $k$-times-iterated natural logarithm (Theorem~\ref{thm:variant3}). We also show that, for a natural class of protocols (including those in this paper), an efficiency of $\Omega\big(1 / (\ln n \cdot \ln\ln n \cdot \ln\ln\ln n \cdots)\big)$, where the product runs over the entire tower of iterated logarithms, is not possible, so the $\epsilon$ in the exponent cannot be removed. Throughout this section, $\ln$ denotes the natural logarithm, while $\log$ continues to denote the base-$2$ logarithm.

The basic idea is to adapt the \prot{Guess-a-Scale} protocol of Section~\ref{sec:basic} to the unknown-$n$ setting. Since the players no longer know $\mathcal{I} = \{0, 1, \ldots, \lfloor \log n \rfloor\}$, they cannot draw $i_t$ uniformly from $\mathcal{I}$. Instead, they draw $i_t$ from a distribution that is, in some sense, as close as possible to being uniform on $\{0, 1, \ldots, \lfloor \log n \rfloor\}$ for all $n$ simultaneously. 

\begin{definition}[The \prot{Guess-a-Scale}$_{r,\epsilon}$ protocol]
\label{def:blind}
Fix an integer $r \ge 1$ and a real $\epsilon > 0$, and let $\ln^{(k)}$ denote the $k$-times iterated natural logarithm, floored at $1$ so that it is defined and at least $1$ everywhere: $\ln^{(1)} x = \max(1, \ln x)$ and $\ln^{(k+1)} x = \max\big(1, \ln(\ln^{(k)} x)\big)$. For each $i \ge 0$, define the weight
\[
  w_i \;=\; \frac{1}{(i+1) \cdot \ln^{(1)}(i+1) \cdots \ln^{(r-1)}(i+1) \cdot \big(\ln^{(r)}(i+1)\big)^{1+\epsilon}},
\]
let $Z = \sum_{i \ge 0} w_i$, and let $q$ be the probability distribution on $\mathbb{Z}_{\ge 0}$ given by $q_i = w_i / Z$. The players use their shared randomness to draw independent scales $i_1, i_2, \ldots$ from $q$, one per step. On step $t$, every active player broadcasts independently with probability $2^{-i_t}$.
\end{definition}

Note that, in the definition above, the quantity $Z = \sum_{i \ge 0} w_i$ is finite (see, e.g., \cite[§4.3, Exercise 11, p.~230]{Trench2013RealAnalysis}), so the distribution $q$ is well-defined.

\begin{restatable}{theorem}{variantthree}
\label{thm:variant3}
Fix a constant integer $r \ge 1$ and a constant $\epsilon > 0$, and suppose the players have a global clock and shared randomness but no knowledge of $n$. Then \prot{Guess-a-Scale}$_{r,\epsilon}$ achieves efficiency
\[
  \Omega\!\left(\frac{1}{\ln n \cdot \ln^{(2)} n \cdots \ln^{(r)} n \cdot \big(\ln^{(r+1)} n\big)^{1+\epsilon}}\right)
\]
against an adaptive scheduler and with no feedback. The hidden constant depends on both $r$ and $\epsilon$.
\end{restatable}

\begin{proof}
The proof of Theorem~\ref{thm:variant1} goes through unchanged except for the probability of guessing the right scale. On a live step with $J$ active players, the good scale is $i^\star = \lfloor \log J \rfloor \le \log n$, and since $q$ is decreasing it is drawn with probability
\begin{align*}
  q_{i^\star} &\;\ge\; q_{\lfloor \log n \rfloor} \\
  &\;=\; \frac{1}{Z} \cdot \frac{1}{x \cdot \ln^{(1)} x \cdots \ln^{(r-1)} x \cdot \big(\ln^{(r)} x\big)^{1+\epsilon}},
\end{align*}
where $x = \lfloor \log n \rfloor + 1$.
Using the fact that $x = \Theta(\ln n)$ we have that $\ln^{(k)} x = \Theta\big(\ln^{(k+1)} n\big)$ for each $k \le r$. Since $r$ and $\epsilon$ are constants, we can conclude that
\[
  q_{\lfloor \log n \rfloor} \;=\; \Omega\!\left(\frac{1}{\ln n \cdot \ln^{(2)} n \cdots \ln^{(r)} n \cdot \big(\ln^{(r+1)} n\big)^{1+\epsilon}}\right).
\]
Using this bound in place of the $\Theta(1 / \log n)$ bound in the proof of Theorem~\ref{thm:variant1}, the result follows from the same arguments as in the proof of Theorem~\ref{thm:variant1}.
\end{proof}

Finally, we conclude the section by proving that there is no single distribution for $i_t$ that can achieve an efficiency of $\Omega\big(1 / (\ln n \cdot \ln\ln n \cdot \ln\ln\ln n \cdots)\big)$ for all $n$ simultaneously, where the product runs over the entire tower of iterated logarithms.
\begin{proposition}
\label{prop:noknowledge}
Consider the protocol of Definition~\ref{def:blind}, but with an arbitrary probability distribution $q$ on $\mathbb{Z}_{\ge 0}$ in place of the one defined there, where $q_i$ denotes the probability that $q$ assigns to $i$. No matter how $q$ is chosen, there are infinitely many values of $n$ for which the efficiency is
\[
  o\!\left(\frac{1}{\ln n \cdot \ln\ln n \cdot \ln\ln\ln n \cdots}\right),
\]
where the product runs over the entire tower of iterated logarithms; that is, the denominator is $\ln^{(1)} n \cdot \ln^{(2)} n \cdot \ln^{(3)} n \cdots$, with the floored iterated logarithms of Definition~\ref{def:blind}, all but finitely many of which equal $1$.
This is true even for the oblivious scheduler that schedules every player on every step.
\end{proposition}

We remark that the use of $\ln$ rather than $\log$ in Proposition \ref{prop:noknowledge} is not just aesthetic. Because there are a super-constant number of logarithms in the denominator, the functions $1 / (\log n \log \log n \cdots)$ and $1 / (\ln n \ln \ln n \cdots)$ are not asymptotically the same. And, whereas the first of the following two series converges, the second diverges:
\[
  \sum_{j \ge 1} \frac{1}{j \log j \log \log j \cdots},
  \qquad
  \sum_{j \ge 1} \frac{1}{j \ln j \ln \ln j \cdots}.
\]

\begin{proof}
Because a fresh scale $i_t \sim q$ is drawn on each step, and because every active player then broadcasts with the same probability $2^{-i_t}$, the probability that a step succeeds depends only on the number of active players. For an integer $a \ge 1$, let $\Phi(a)$ denote that probability when exactly $a$ players are active, and for $N \ge 0$ let
\[
  M_N \;=\; \max\big\{ \Phi(a) \;:\; 2^{N} < a \le 2^{N+1} \big\}
\]
be the largest of these probabilities as $a$ ranges over the values between $2^N$ and $2^{N+1}$. The main step of the proof is to show that 
\begin{equation}
\label{eq:sum-MN}
  \sum_{N \ge 0} M_N \;=\; O(1).
\end{equation}

To prove \eqref{eq:sum-MN}, fix a number of players $a$ and a scale $i$ and let $\delta = \log a - i$, so that $2^{\delta} = a 2^{-i}$ is the expected number of broadcasters when the scale is $i$. By the one-success formula, $\Phi(a) = \sum_{i \ge 0} q_i\, h(a,i)$ where $h(a,i) = a\, 2^{-i}(1 - 2^{-i})^{a-1}$, and we claim that there is an absolute constant $C$ satisfying $h(a,i) \le C\,2^{-|\delta|}$ for all $a$ and $i$. If $\delta \le 0$, this is immediate from $h(a,i) \le a2^{-i} = 2^{\delta}$. If $\delta > 0$, write $\lambda = 2^{\delta} \ge 1$ and use $1 - x \le e^{-x}$ along with $2^{-i} \le 1$ to get
\begin{align*}
  h(a,i) &\;\le\; \lambda e^{-(a-1)2^{-i}} \;\le\; \lambda e^{-(\lambda - 1)} \;=\; e \cdot \lambda e^{-\lambda} \\
  &\;=\; O(1/\lambda) \;=\; O\big(2^{-|\delta|}\big),
\end{align*}
where the second-to-last step uses the fact that $\lambda^2 e^{-\lambda}$ is bounded.

Now fix $N$ and let $a$ be any number of players in the range $2^N < a \le 2^{N+1}$. Then $N \le \log a \le N+1$, so $|\delta| = |\log a - i| \ge |N - i| - 1$, and therefore
\[
  h(a,i) \;\le\; C\,2^{-|\delta|} \;\le\; C\,2^{-|N-i|+1} \qquad \text{for every $i \ge 0$.}
\]
Since this holds for every such $a$, it holds in particular for the $a$ that maximizes $\Phi(a)$ over the range, and so
\[
  M_N \;=\; \Phi(a) \;=\; \sum_{i \ge 0} q_i\, h(a,i) \;\le\; 2C \sum_{i \ge 0} q_i\, 2^{-|N-i|}.
\]
Summing this bound over $N$, and then exchanging the order of the two summations,
\begin{align*}
  \sum_{N \ge 0} M_N &\;\le\; 2C \sum_{N \ge 0} \sum_{i \ge 0} q_i\, 2^{-|N-i|}
  \;=\; 2C \sum_{i \ge 0} q_i \sum_{N \ge 0} 2^{-|N-i|} \\
  &\;\le\; 6C \sum_{i \ge 0} q_i \;=\; 6C,
\end{align*}
where we have used that $\sum_{N \ge 0} 2^{-|N-i|} \le \sum_{d \in \mathbb{Z}} 2^{-|d|} = 3$ for every $i$, and that $q$ is a probability distribution. This proves \eqref{eq:sum-MN}.

We can now use \eqref{eq:sum-MN} to find values of $N$ for which $M_N$ is small. For $N \ge 0$, let
\[
  g(N) \;=\; \ln^{(1)} \big(2^{N+1}\big) \cdot \ln^{(2)} \big(2^{N+1}\big) \cdot \ln^{(3)} \big(2^{N+1}\big) \cdots
\]
be the product of the entire tower of iterated logarithms of $2^{N+1}$. We use the standard fact that the series
\[
  \sum_{m \ge 1} \frac{1}{m \cdot \ln^{(1)} m \cdot \ln^{(2)} m \cdot \ln^{(3)} m \cdots}
\]
diverges, where again the product runs over the entire tower. Writing $y = \ln^{(1)}\big(2^{N+1}\big) = \max\big(1, (N+1)\ln 2\big) \le N + 1$, and using $\ln^{(k+1)} x = \ln^{(k)}\big(\ln^{(1)} x\big)$, we have $g(N) = y \cdot \ln^{(1)} y \cdot \ln^{(2)} y \cdots \le (N+1) \cdot \ln^{(1)}(N+1) \cdot \ln^{(2)}(N+1) \cdots$, since each $\ln^{(k)}$ is nondecreasing. Hence $\sum_{N \ge 0} 1/g(N)$ diverges as well. Suppose that, for some $c > 0$, we had $M_N \ge c/g(N)$ for all sufficiently large $N$. Then $\sum_{N \ge 0} M_N \ge c \sum_{N \ge N_0} 1/g(N)$ for some $N_0$, and the right-hand side diverges, contradicting \eqref{eq:sum-MN}. So no such $c$ exists, which is to say that there is an infinite set $\mathcal{N}$ of values of $N$ along which $g(N) \cdot M_N \to 0$. Fix $N \in \mathcal{N}$, take $n = 2^{N+1}$ players, all arriving on step $1$, and let the scheduler schedule every player on every step. Every step is then live until the last player leaves, and whenever exactly $a$ players are present, each step succeeds with probability exactly $\Phi(a)$, whatever the history is. The number of steps on which exactly $a$ players are present is therefore geometric with mean $1/\Phi(a)$, and counting only the steps on which more than $n/2$ players are present,
\[
  \E[L] \;\ge\; \sum_{a = 2^N + 1}^{2^{N+1}} \frac{1}{\Phi(a)} \;\ge\; \frac{2^N}{M_N} \;=\; \frac{n}{2 M_N}.
\]
The efficiency is thus at most $2 M_N$. Since $g(N) = \ln^{(1)} n \cdot \ln^{(2)} n \cdot \ln^{(3)} n \cdots$ and $g(N) \cdot M_N \to 0$ along $\mathcal{N}$, the efficiency is $o\big(1/(\ln n \cdot \ln\ln n \cdot \ln\ln\ln n \cdots)\big)$.
\end{proof}

\section{The Importance of Implicit Shared State}
\label{sec:shared}
The two resources that the players in Sections~\ref{sec:basic} through~\ref{sec:unknownn} rely on -- a global clock and shared randomness -- can both be viewed as forms of implicit shared state. In this section, we show that such state is fundamentally necessary to get good efficiency. Without it, no protocol achieves efficiency better than $O(\log n / n)$ (Theorem~\ref{thm:lb-noclock}). As noted in the following proposition, this is within a logarithmic factor of the $\Omega(1/n)$ bound achieved by the trivial broadcast-with-probability $1/n$ protocol.

\begin{proposition}[The trivial protocol]
\label{prop:trivial}
Suppose the players know $n$. If every active player broadcasts with probability $1/n$ on every step, then $\E[L] \le 4n^2$, so the efficiency is $\Omega(1/n)$. This does not require any clock, shared randomness, or feedback, and it holds against any adaptive scheduler.
\end{proposition}
 
\begin{proof}
On a live step with $J \ge 1$ active players, the one-success formula gives success probability $\frac Jn (1 - \frac1n)^{J-1} \ge \frac{1}{n}(1-\frac1n)^{n-1} \ge \frac{1}{en} \ge \frac{1}{4n}$. This holds no matter what the history is and no matter which players the scheduler schedules, so Lemma~\ref{lem:accounting-ub} applies with $q = 1/(4n)$ to give $\E[L] \le n/q = 4n^2$.
\end{proof}

The rest of the section is spent proving the following theorem.

\lbnoclock*

\begin{proof}[Proof of Theorem~\ref{thm:lb-noclock}]
Fix any player protocol. The main challenge in the proof is to construct the scheduler that we will use against it. The scheduler runs in \defn{rounds}, where each round is either:
\begin{itemize}[itemsep=2pt]
  \item a \defn{broad round}, which is a single step on which every player is scheduled;
  \item a \defn{narrow round}, which is a block of $n$ consecutive steps, the $i$-th of which schedules player $i$ alone.
\end{itemize}
Note that, when deciding whether to broadcast during a given round, the players cannot tell what type of round is occurring. In both cases, each player gets exactly one turn. Since players do not have a global clock, they experience only their own turns, so after $r$ rounds every player has lived through exactly $r$ turns, no matter which types of round those were.

Our scheduler is oblivious, so the type of each round is decided before the execution begins. Write $\tau_r \in \{\mathrm{broad}, \mathrm{narrow}\}$ for the type of round $r$, and call the sequence $\tau = (\tau_1, \tau_2, \ldots)$ the \defn{schedule}.

\medskip
\noindent\textbf{The ghost process.} To choose whether a round should be broad or narrow, it will be helpful for the scheduler to reason about a fictitious process in which broad rounds always produce collision failures. In the \defn{ghost process} for a schedule $\tau$, everything is as in the real execution of $\tau$ except that, regardless of how many players actually broadcast, the players are told that the round failed due to a collision; and even if exactly one player broadcasts, that player remains present. Narrow rounds behave normally.

The key property that makes the ghost process useful is that, for every schedule, the players do not interact:

\begin{claim}
\label{clm:ghost-indep}
Let $\tau$ be any schedule, and consider the ghost process for $\tau$. For each player $i$ and each round $r$, let $X_{i,r}$ be the indicator random variable for the event that player $i$ is present at round $r$ and broadcasts on its turn in round $r$, and let $X_i = (X_{i,1}, X_{i,2}, \ldots)$. Then the sequences $X_1, \ldots, X_n$ are mutually independent. In particular, for every round $r$, the indicators $X_{1,r}, \ldots, X_{n,r}$ are mutually independent.
\end{claim}

\begin{proof}
Since the players have no global clock, the only information that player $i$ ever receives is the feedback from the rounds that it takes part in, and it takes part in exactly one turn per round. On a broad round, the ghost process reports a collision failure to $i$, no matter what any of the players actually did. On a narrow round, $i$ is the only scheduled player, so the player either broadcasts and succeeds or does not broadcast and hears silence. From player $i$'s perspective, then, the game being played is this: on each round, $i$ gets to choose whether to broadcast or not; it then learns whether the round was broad or narrow, and it succeeds if and only if the round was narrow and it chose to broadcast. 

Since there is no shared randomness, since $\tau$ is fixed (not random), and since there is no interaction between the games being played by different players, the only randomness in this game is the players' own private random bits. These random bits are mutually independent across players, and hence so are the sequences $X_1, \ldots, X_n$.
\end{proof}

\medskip
\noindent\textbf{The scheduler.} Having defined the ghost process, we can now describe how the scheduler chooses which rounds should be broad or narrow. 

Let $\theta = 40 \ln n$. The scheduler determines $\tau_1, \tau_2, \ldots$ one after another, as follows. Suppose that $\tau_1, \ldots, \tau_{r-1}$ have already been determined, and let $X_{i,r}$ be as in Claim~\ref{clm:ghost-indep} (critically, the $X_{i, r}$s are determined by the \emph{ghost process}, and so later we will be able to use the fact that they are independent across $i$). The distribution of $X_{i,r}$ is completely determined by $\tau_1, \ldots, \tau_{r-1}$ (and by the description of the protocol that the players are using, which the scheduler can be assumed to know). So the number
\[
  \mu_r \;=\; \E\Big[\sum_{i} X_{i,r}\Big],
\]
which we can think of as being the \emph{contention} in round $r$, is also completely determined by $\tau_1, \ldots, \tau_{r-1}$. Thus, even though the scheduler is oblivious, it can compute $\mu_r$. The scheduler sets $\tau_r = \mathrm{broad}$ if $\mu_r > \theta$, and $\tau_r = \mathrm{narrow}$ otherwise. This completes the description of the scheduler.

For the rest of the proof, $\tau$ denotes the schedule that we have just constructed, and $X_{i,r}$ (along with the sequences $X_i$) denote the indicators defined in Claim~\ref{clm:ghost-indep} for that schedule.

\medskip
\noindent\textbf{The ghost process is faithful.} To analyze the scheduler, we first observe that, with good probability, the real and ghost processes agree throughout the first $n^3$ rounds.
Couple the real and ghost processes by giving them the same player random bits. They agree step for step until the first broad round in which fewer than two present players broadcast, since that is the only place where their rules differ. Consider a broad round $r$, so that $\mu_r > \theta$. The number of players that broadcast in round $r$ is $\sum_i X_{i,r}$, which by Claim~\ref{clm:ghost-indep} is a sum of independent indicators with mean $\mu_r > \theta$. So, by the standard lower-tail Chernoff bound,
\begin{align*}
  \Prb[\,\text{at most one broadcaster}\,] &\;\le\; \Prb[\,\#\text{broadcasters} \le \theta/2 \,] \\
  &\;\le\; e^{-\theta/8} \;=\; e^{-5\ln n} = n^{-5},
\end{align*}
where the first inequality uses $\theta/2 \ge 1$. Union bounding over the first $n^3$ rounds, the real and ghost processes agree throughout those rounds with probability at least $1 - n^{-2}$. For the rest of the proof, we consider only the first $n^3$ rounds, we work with the ghost process, and we add $n^{-2}$ to the failure probability at the end.

\medskip
\noindent\textbf{Analyzing the efficiency.} Let $m = \lfloor n / (320 \ln n)\rfloor$, and assume $n$ is large enough that $m \ge n/(640 \ln n) \ge 1$. In a narrow round $r$, each player that broadcasts is the only scheduled player when it does so, so the number of successes is exactly $\sum_i X_{i,r}$, whose expectation is $\mu_r \le \theta$. Consider the first $\min(m, m^\star)$ narrow rounds, where $m^\star$ is the number of narrow rounds among the first $n^3$. The total number of successes across these rounds is $\sum_i \ind[i \text{ succeeds in one of them}]$, and each term here is determined by $X_i$, so by Claim~\ref{clm:ghost-indep} this is again a sum of independent indicators. Its mean is at most $m\theta \le n/8$. By the standard upper-tail Chernoff bound, the probability that it reaches $n/4$ is at most $e^{-n/24}$.

So, except with probability $e^{-n/24}$, fewer than $n/4$ players have left the system during those rounds, i.e.\ at least $3n/4$ players are present throughout them. Now split into two cases.

If $m^\star \ge m$, then each of the first $m$ narrow rounds contains at least $3n/4$ live steps (one per surviving player), so
\[
  L \;\ge\; m \cdot \frac{3n}{4} \;=\; \Omega\!\left(\frac{n^2}{\log n}\right).
\]
If instead $m^\star < m$, then at least $n^3 - m$ of the first $n^3$ rounds are broad rounds, and each of them is a live step because at least $3n/4 \ge 1$ players are present; so $L \ge n^3 - m \ge n^2$, which is even larger.

Thus, with probability at least $1 - n^{-2} - e^{-n/24}$ we have $L = \Omega(n^2/\log n)$, and therefore $\E[L] = \Omega(n^2/\log n)$. The efficiency is $n/\E[L] = O(\log n/n)$, which completes the proof.
\end{proof}

\section*{Acknowledgments}

Part of this research was conducted while attending the Circles Workshop on Concurrent Data Structures at the Jane Street office in New York, supported through the Circles, the Simons Institute – Jane Street Small Group Collaborations program.

 M.~A.~Bender was supported by NSF grants CCF 2247577,
IIS 2551616, Broadcom, Sandia National Laboratories, and the John L. Hennessy Chaired Professorship. P. Kundu and M. Kuszmaul were supported by NSF grants CCF 2247577 and IIS 2551616. M. Farach-Colton was supported by NSF Grants 2551614, 2420942, 2247576, and 2230742.
A. Krapivin supported by the Jeanne~B.\ and Richard~F.\ Berdik ARCS Pittsburgh Endowed Scholar Award, NSF grants CNS-2504471 and CCF-2542165, a Jane Street Research Grant, the David and Lucile Packard Foundation (Award \#202071730), and gifts from Nikolai Mushegian and Microsoft. J. Kuszmaul was supported by NSF Grant CCF-2330048, BSF Grant 2024233, a Simons
Investigator Award, and an MIT Akamai Presidential Fellowship.
W.~Kuszmaul and J.~Liang were supported by NSF grants CNS-2504471 and CCF-2542165, and a Jane Street Research Grant.

\subsection*{AI Acknowledgments}All of the technical results in this paper are human generated. AI tools (GPT 5.6, GPT 6, Opus 5.0, Opus 5.5) were used to assist in writing and editing the paper, and for filling in routine calculations. All portions generated by AI were then heavily edited by the authors.

\bibliographystyle{ACM-Reference-Format}
\bibliography{references}

\appendix

\section{Coarse-Grained Clocks Against an Adaptive Scheduler}
\label{app:adaptive-coarse}
In Section~\ref{sec:coarse}, we showed that, against an oblivious scheduler, a $k$-granular clock costs the players nothing so long as $k \in O(\log n)$ (Theorem~\ref{thm:variant2}). In this appendix, we show that the assumption of an oblivious scheduler is necessary: against an adaptive scheduler, a $k$-granular clock costs the players a factor of $k$, even if they receive ternary feedback and complete information about all previous windows. As in Section~\ref{sec:coarse}, for $w = 1, 2, 3, \ldots$, we refer to the block of steps $\{(w-1)k + 1, \ldots, wk\}$ on which the clock reads $w$ as the \defn{window} $w$, and to step $(w-1)k + i$ as the \defn{$i$-th step} of window $w$.

\begin{proposition}
\label{prop:adaptive-coarse-lb}
Suppose that $1 \le k \le n^{1/4}$. Then there exists an adaptive scheduler against which every player protocol satisfies $\E[L] = \Omega(kn\log n)$ and hence has efficiency $O(1/(k \log n))$. This holds even when the players have a $k$-granular clock, shared randomness, ternary feedback, and knowledge of $n$. In fact, it holds even if every player that is scheduled during a window is given complete information about all previous windows (i.e., for each step of each previous window, which players were scheduled and which ones broadcast).
\end{proposition}

Note that the lemma is tight. Indeed, if the players run the \prot{Guess-a-Scale} protocol from Section~\ref{sec:basic} with one shared scale per window rather than per step, then the scale is still fresh on the first live step of each window. The proof of Theorem~\ref{thm:variant1} shows that this step succeeds with probability $\Omega(1/\log n)$. And, since each window contains at most $k$ live steps, it follows that $\E[L] = O(kn\log n)$, even against an adaptive scheduler.

The high-level approach that the scheduler takes for Proposition \ref{prop:adaptive-coarse-lb} is as follows. What the scheduler achieves is that, in each window, all steps are live, and the expected number of successes is $O(1/\log n)$ (at least while $\geq n/2$ players are present). The first step of each window is a \emph{probe}: similarly to the scheduler in the proof of Theorem~\ref{thm:lb-clock}, the scheduler guesses a random scale and samples a random number of players according to the scale. As in the earlier proof, players can succeed on the probe step with probability $O(1/\log n)$ at best, but now the probe also teaches the scheduler roughly how likely the players are to broadcast. If the probe is silent or successful, then few players are willing to broadcast, so on the remaining $k - 1$ steps of the window, the scheduler schedules one player at a time, and is unlikely to choose any broadcasting players. If the probe is noisy (i.e., results in a collision), then many players are willing to broadcast, so the scheduler instead schedules large groups of players at a time, which are likely to collide. Crucially, the scheduler never schedules any player twice during the same window. Since the clock reads the same value throughout the window, a player therefore cannot tell whether it is being scheduled on the probe step or on one of the later steps and thus cannot change their own behavior in response to the scheduler. This intuition is formalized in the following proof.

\begin{proof}[Proof of Proposition~\ref{prop:adaptive-coarse-lb}]
Throughout the proof, we assume that $n$ is sufficiently large (otherwise, the lemma follows directly from Theorem~\ref{thm:lb-clock}). Define
\begin{gather*}
  r_{\min} = k n^{1/8}, \qquad
  r_{\max} = \frac{n^{3/4}}{k}, \\
  \mathcal{J} = \{ j \in \mathbb{Z} : r_{\min} \le 2^j \le r_{\max} \}.
\end{gather*}
Since $k^2 \le n^{1/2}$, the interval $[\log r_{\min}, \log r_{\max}]$ has length $\frac58 \log n - 2\log k \ge \frac18 \log n$, so
\[
  |\mathcal{J}| \;\ge\; \frac18 \log n - 1 \;=\; \Omega(\log n).
\]
We start by defining the scheduler and then analyze its effect on any protocol.

\medskip
\noindent\textbf{The scheduler.} Consider a window $w$, and let $P$ be the set of players that are present at the start of the window.
If $|P| < n/2$, the scheduler simply schedules all of $P$ on every step of the window (such windows will play no role in the analysis). Otherwise, using fresh randomness, 
the scheduler acts as follows.
\begin{itemize}[itemsep=2pt]
  \item \textbf{The probe step (step $1$ of the window).} The scheduler chooses a uniformly random $j \in \mathcal{J}$, sets $r = 2^j$, and schedules a uniformly random set $S \subseteq P$ of exactly $r$ players (note $|S| \leq r_{\max} \leq n/4$).
  \item \textbf{The remaining steps (steps $2, \ldots, k$ of the window).} All players scheduled on these steps are drawn from $P \setminus S$, and no player is scheduled on more than one of them. There are two cases.
  \begin{itemize}[itemsep=2pt]
    \item \defn{Quiet case} (the probe step was silent or successful). The scheduler chooses $k - 1$ distinct players uniformly at random from $P \setminus S$, and schedules the $i$-th of them alone on step $i + 1$ of the window.
    \item \defn{Loud case} (the probe step was noisy). The scheduler partitions $P \setminus S$ uniformly at random into $k - 1$ groups whose sizes differ by at most one, and schedules the $i$-th group on step $i + 1$ of the window.
  \end{itemize}
\end{itemize}
\medskip
\noindent\textbf{Every step is live.} First note that every step is guaranteed to be live. If $|P| < n/2$, the scheduler schedules all of $P$ on every step, so every step is live. Otherwise, as $n$ is sufficiently large, at least one player is scheduled on every step, and, as every step schedules players that were present at the end of the previous window and that were not yet scheduled in the current window, every scheduled player is present.

\medskip
\noindent\textbf{Players cannot tell the steps of a window apart.} Fix a window $w$, and condition on the history through the end of window $w - 1$ taking some particular value $h$ in which at least $n/2$ players are present, \emph{and on all the randomness that the players will use during window $w$}. Consider a player $p \in P$. The scheduler schedules $p$ at most once during window $w$, and throughout the window the clock reads $w$. So, whenever $p$ is scheduled during window $w$, it has not yet received any feedback during the window, and everything it knows is determined by what we have conditioned on. It follows that $p$'s decision whether to broadcast if scheduled during window $w$ is fixed, and, critically, is the same no matter which step of the window $p$ is scheduled on.

We define $A$ to be the set of players in $P$ that would broadcast if scheduled during the window, and let $a = |A|$. Under the conditioning above, $A$ and $a$ are fixed, and the only remaining randomness is the scheduler's randomness in window $w$. Since every scheduled player is present, a step of the window succeeds if and only if exactly one player of $A$ is scheduled on it. In particular, letting $b = |S \cap A|$, the probe step is silent or successful if $b \le 1$ and noisy if $b \ge 2$.

\medskip
\noindent\textbf{A sampling fact.} Both the probe set and each group in the loud case are uniformly random sets of a fixed size, so we will analyze them using the following standard fact. Let $U$ be a finite set, let $B \subseteq U$ be fixed, let $T$ be a uniformly random subset of $U$ of some fixed size, and let $X = |T \cap B|$ and $\mu = \E[X] = |B|\,|T|/|U|$. Then
\begin{align}
  \Prb[\, X \ge t \,] &\;\le\; \Big(\frac{e\mu}{t}\Big)^{t} \qquad \text{for every } t \ge \mu, \label{eq:sample-upper-tail} \\
  \Prb[\, X \le \mu/2 \,] &\;\le\; e^{-\mu/8}. \label{eq:sample-chernoff}
\end{align}
Both are Chernoff bounds: the indicators $\ind[u \in T]$ for $u \in B$ are negatively associated, since the indicators of membership in a uniformly random set of fixed size are negatively associated~\cite{JoagDevPr83}, and any subset of a family of negatively associated random variables is again negatively associated; and Chernoff bounds apply to sums of negatively associated indicators just as they do to sums of independent ones~\cite{DubhashiRa98}.

\medskip
\noindent\textbf{What the probe reveals.} Roughly, the following claim says that, with high probability, if $a$ is large, the probe will be noisy, and, if $a$ is small, the probe is quiet. Let
\[
  a_{\mathrm{lo}} = k n^{1/16}, \qquad a_{\mathrm{hi}} = \frac{n^{15/16}}{k}.
\]

\begin{claim}
\label{clm:probe}
Condition on everything above. Then, whatever the scale $j$ is:
\begin{enumerate}[label=(\alph*)]
  \item If $a < a_{\mathrm{lo}}$, then the probe step is noisy with probability at most $8n^{-3/8}$.
  \item If $a \ge a_{\mathrm{lo}}$, then $\Prb[\, b > a/2 \,] \le n^{-2}$.
  \item If $a \ge a_{\mathrm{hi}}$, then the probe step is silent or successful with probability at most $n^{-2}$.
\end{enumerate}
\end{claim}

\begin{proof}
Fix the scale $j$, so that $r = 2^j \in [r_{\min}, r_{\max}]$, and let $m = |P|$, so that $n/2 \le m \le n$. We apply the sampling fact with $U = P$, $B = A$, and $T = S$, so that $X = b$ and $\mu = ar/m$.

Parts (a) and (b) use only that the probe set is small: $r \le r_{\max}$. For part (a), this gives $\mu \le 2a_{\mathrm{lo}}\, r_{\max}/n = 2n^{-3/16}$, so~\eqref{eq:sample-upper-tail} with $t = 2$ gives $\Prb[b \ge 2] \le (e\mu/2)^2 \le e^2 n^{-3/8} \le 8n^{-3/8}$. For part (b), this gives $\mu = ar/m \le 2ar_{\max}/n \le 2a n^{-1/4} \le a/2$, so~\eqref{eq:sample-upper-tail} with $t = a/2$ gives
\[
  \Prb[\, b > a/2 \,] \;\le\; \Big(\frac{2e\mu}{a}\Big)^{a/2} \;\le\; \big(4e\, n^{-1/4}\big)^{a/2} \;\le\; n^{-a/16} \;\le\; n^{-2},
\]
where the last step uses $a \ge a_{\mathrm{lo}} \ge 32$.

Part (c) uses only that the probe set is large: $r \ge r_{\min}$, which gives $\mu \ge a_{\mathrm{hi}}\, r_{\min}/n = n^{1/16} \ge 2$ (using $m \le n$). So~\eqref{eq:sample-chernoff} gives $\Prb[b \le 1] \le \Prb[b \le \mu/2] \le e^{-n^{1/16}/8} \le n^{-2}$.
\end{proof}

\medskip
\noindent\textbf{Successes are rare.} Let $Z_w$ denote the number of successes during window $w$. We bound the expected contributions to $Z_w$ from the probe step, from the quiet case, and from the loud case.

\emph{The probe step.} The analysis of the probe step is similar to that of a step in the proof of Theorem~\ref{thm:lb-clock}: the probe step succeeds only if $b = 1$, which is unlikely unless $\mu = \Theta(1)$, and only $O(1)$ scales give $\mu = \Theta(1)$. Formally, for a fixed scale $j$, we have $\Prb[b = 1] \le \Prb[b \ge 1] \le \mu$, where $\mu = a 2^j/|P|$, and if $\mu \ge 2$, then~\eqref{eq:sample-chernoff} gives $\Prb[b = 1] \le \Prb[b \le \mu/2] \le e^{-\mu/8}$. As $j$ ranges over $\mathcal{J}$, the value of $\mu$ doubles from each scale to the next. Thus, since $j$ is uniform over $\mathcal{J}$, the scales with $\mu < 2$ contribute at most
\[
  \frac{1}{|\mathcal{J}|}\Big(2 + 1 + \frac12 + \cdots\Big) \;=\; O(1/\log n)
\]
to the probability that the probe step succeeds, and the scales with $\mu \ge 2$ contribute at most
\[
  \frac{1}{|\mathcal{J}|} \sum_{i \ge 0} e^{-2^{i+1}/8} \;=\; O(1/\log n),
\]
for a total success probability of $O(1/\log n)$.

\emph{The quiet case.} If $a \ge a_{\mathrm{hi}}$, then by Claim~\ref{clm:probe}(c), the quiet case occurs with probability at most $n^{-2}$, so it contributes at most $k n^{-2}$ expected successes. Otherwise, $a < a_{\mathrm{hi}}$. Each of steps $2, \ldots, k$ then schedules a single player, uniformly random from the at least $n/4$ players of $P \setminus S$, so each of these steps succeeds with probability at most $4a/n < 4a_{\mathrm{hi}}/n$. The quiet case then contributes at most
\[
  (k - 1) \cdot \frac{4 a_{\mathrm{hi}}}{n} \;\le\; 4 n^{-1/16}
\]
expected successes. Intuitively, if $a$ is large, then the probe is unlikely to be quiet; and if $a$ is small, then a randomly chosen player is unlikely to lie in $A$.

\emph{The loud case.} If $a < a_{\mathrm{lo}}$, then by Claim~\ref{clm:probe}(a), the loud case occurs with probability at most $8n^{-3/8}$, so it contributes at most $8k n^{-3/8} \le 8n^{-1/8}$ expected successes. Otherwise, $a \ge a_{\mathrm{lo}}$. By Claim~\ref{clm:probe}(b), the event $b > a/2$ occurs with probability at most $n^{-2}$ and so contributes at most $k n^{-2}$ expected successes. So condition on a probe set $S$ with $b \ge 2$ and $b \le a/2$, so that at least $a/2$ players of $A$ lie in $P \setminus S$. By linearity of expectation, it suffices to bound the success probability of each of steps $2, \ldots, k$ individually. Consider step $i + 1$, and let $G$ be the $i$-th group, which on its own is a uniformly random subset of $P \setminus S$ of size at least $\lfloor |P \setminus S|/(k-1) \rfloor \ge |P \setminus S|/(2k)$. We apply the sampling fact with $U = P \setminus S$, $B = A \setminus S$, and $T = G$, so that $X = |G \cap A|$ and
\[
  \mu \;\ge\; \frac{a}{2} \cdot \frac{1}{2k} \;\ge\; \frac{a_{\mathrm{lo}}}{4k} \;=\; \frac14 n^{1/16}.
\]
The step succeeds only if $X = 1$, so, by~\eqref{eq:sample-chernoff}, it succeeds with probability at most $\Prb[X \le \mu/2] \le e^{-n^{1/16}/32} \le n^{-2}$. Summing over the $k - 1$ steps, the loud case contributes at most $2k n^{-2} \le 8n^{-1/8}$ expected successes in this case, as well. Intuitively, if $a$ is small, then the probe is unlikely to be noisy; and if $a$ is large, then every group contains many players of $A$, and so collides.

\emph{Putting the pieces together.} The bounds for the quiet and loud cases hold for every scale $j$, while the bound for the probe step holds after averaging over $j$. Summing the three, and conditioning on the history $h$ and on the players' randomness (but no longer on $j$),
\[
  \E[\, Z_w \,] \;\le\; O(1/\log n) + 4n^{-1/16} + 8n^{-1/8} \;=\; O(1/\log n).
\]
Since this bound holds for every value of the players' randomness during window $w$, it also holds after averaging over that randomness, i.e., conditioned only on the history $h$.

\medskip
\noindent\textbf{Combining the bounds.} We finish with an argument similar to the proof of Lemma~\ref{lem:accounting-lb}, but with windows in place of steps. Let $q = O(1/\log n)$ be the bound on $\E[Z_w]$ from above, and let $W$ be the number of windows that start with at most $n/2$ successes so far (i.e., with at least $n/2$ players present). Since the number of successes never decreases, these are exactly windows $1, \ldots, W$. Since every step is live, $L \ge kW$, so it suffices to show that $\E[W] \ge n/(2q)$.

If $W = \infty$ with positive probability, this is immediate. Otherwise, $W$ is finite with probability $1$, and, since window $W + 1$ starts with more than $n/2$ successes, more than $n/2$ successes occur during windows $1, \ldots, W$. The event $w \le W$ is determined by the history through the end of window $w - 1$, so the bound on successes from above gives
\begin{align*}
  \frac n2 \;\le\; \E\Big[\sum_{w = 1}^{W} Z_w\Big]
  &\;=\; \sum_{w \ge 1} \E\big[Z_w \cdot \ind[w \le W]\big] \\
  &\;\le\; q \sum_{w \ge 1} \Prb[w \le W] \;=\; q\, \E[W].
\end{align*}
Therefore,
\[
  \E[L] \;\ge\; k\, \E[W] \;\ge\; k \cdot \frac{n}{2q} \;=\; \Omega(kn \log n). \qedhere
\]
\end{proof}

\end{document}